\input amstex
\magnification=1200
\documentstyle{amsppt}
\NoRunningHeads \loadbold \pagewidth{135 mm} \topmatter
\title
Darboux transformation for classical acoustic spectral problem
\endtitle
\author
A.A. Yurova*, A.V. Yurov$\dag$, M. Rudnev$\ddag$
\endauthor
\address *Department of Mathematics, Kaliningrad State Technical
University, Sovetsky Pr. 1, Kaliningrad 236000, Russia\newline\indent
$\dag$Department of Theoretical Physics, Kaliningrad State University, 236041,
Aleksandra Nevskogo 14, Kaliningrad 236041, Russia\newline\indent $\ddag$
School of Mathematics, University Walk, Bristol BS8 1TW, UK
\endaddress
\email artyom\_yurov\@mail.ru; M.Rudnev\@bris.ac.uk
\endemail
\thanks This work was supported by RFBR Grant 00-01-00783,
the Grant of Education Department of the Russian Federation,  No. E00-3.1-383,
and the NSF Grant DMS 0072153.
\endthanks

\abstract We study discrete isospectral symmetries for the classical acoustic
spectral problem in spatial dimensions one and two, by developing a Darboux
(Moutard) transformation formalism for this problem. The procedure follows the
steps, similar to those for the Schr\"{o}dinger operator. However, there is no
one-to-one correspondence between the two problems. The technique developed
enables one to construct new families of integrable potentials for the acoustic
problem, in addition to those already known.

The acoustic problem produces a non-linear Harry Dym PDE. Using the technique,
we reproduce a pair of simple soliton solutions of this equation. These
solutions are further used to construct a new positon solution for this PDE.
Furthermore, using the dressing chain approach, we build a modified Harry Dym
equation together with its LA-pair.

As an application, we construct some singular and non-singular integrable
potentials (dielectric permitivity) for the Maxwell equations in a 2D
inhomogeneous medium.

\endabstract
\endtopmatter
\document
\head Introduction \endhead

This note develops the Darboux transformation and dressing chain formalism for
the classical acoustic spectral problem (below just the ``acoustic problem'')
and the related Harry Dym (HD) equation. It treats the problem in the same vein
as it is done for the Schr\"{o}dinger operator and the related  KdV (mKdV)
hierarchies. The acoustic problem and the  Schr\"{o}dinger operator are closely
connected. This connection constituted a base for the approach to the acoustic
problem and the HD equation in [1-3]. However, as is discussed below, the
relation between the problems is not utterly straightforward.

The acoustic problem describes wave propagation in inhomogeneous acoustic or
electromagnetic media and just like the Schr\"{o}dinger equation is
non-integrable for an arbitrary potential. For applications, it is important to
be able to construct intergable potentials, which result in solutions with
given properties or asymptotic behavior. For instance, for the purposes of
transmission of information, reflexionless potentials are important. These
potentials are such that the problem admits solutions, which asymptote to
$e^{-ipx}$ as $x\rightarrow - \infty$ and  $T(p)e^{ipx}$ as $x\rightarrow
\infty$, with the passage coefficient $T(p)\in\Bbb C$ being one in absolute
value. This was recently studied in a work by Novikov [4], which has drawn our
attention to the problem.

The latter work constructs a family of so-called B-potentials for the acoustic
problem via a semi-classical solution ansatz. We show that these potentials
naturally come up as a result of a one-step Darboux transform on the
``vacuum background''. In continuous media electrodynamics,  potentials of
the acoustic problem can be interpreted as the medium's dielectric permitivity.
The B-potentials in question possess point singularities, and hence
their physical meaning is not entirely clear. On the other hand, the dressing
technique enables one to construct a variety of potentials, which are neither
B-potentials, nor singular. We illustrate it by a single act of dressing chain
closing (in dimension one and two) which yields regular integrable potentials
(dielectric permitivity).

The acoustic problem on the real line is described by the following ODE: $$
\psi_{xx}=\frac{\lambda}{u^2(x)}\psi. \eqno(1) $$ This equation models wave
propagation in non-homogeneous (acoustic or electromagnetic) media.
Consider for instance the Maxwell equations in a medium without external
sources with the standard notations $(\bold E,\bold H)$ for the electromagnetic
field, as well as $\bold D=\epsilon\bold E,\,\bold B=\kappa\bold H$. Suppose,
the medium is isotropic but inhomogeneous with $\kappa\equiv 1$ and $\epsilon
=\epsilon(x,y,z)$. Then

$$ {\text {rot}}\bold B=\frac{1}{c}\frac{\partial\bold D}{\partial t},\qquad
{\text {rot}}\bold E=-\frac{1}{c}\frac{\partial\bold B}{\partial t},\qquad
{\text {div}}\bold D={\text {div}}\bold B=0. \eqno(2) $$

One can easily exclude the quantity $\bold B$ from (2) to obtain an equation
connecting the quantities $\bold E$ and $\bold D$: $$ {\text {rot\,rot}}\bold
E=-\frac{1}{c^2} \frac{\partial^2\bold D}{\partial t^2}. \eqno(3) $$ For the
electric field $\bold E$ one writes $\bold E=e^{i\omega t}\boldsymbol\psi$,
with $\boldsymbol\psi=\boldsymbol\psi(x,y,z)$ and taking into account the last
equation of (2) obtains: $$
\nabla\left(\frac{(\boldsymbol\psi,\nabla)\epsilon}{\epsilon}\right)+
\Delta\boldsymbol\psi=-\frac{\omega^2}{c^2}\epsilon\boldsymbol\psi, \eqno(4) $$
where $\Delta$ is a three dimensional Laplassian.

The equation (1) follows if one lets $\epsilon=\epsilon(x),\,
\boldsymbol\psi=(0,0,\psi(x)),\,\lambda=-\omega^2/c^2,\,u^{-2}(x)=\epsilon(x)$.
The dielectric permitivity $\epsilon(x)$ will be henceforth referred to as a
potential.

Alternatively, one can choose $\epsilon=\epsilon(x,y)$ as well as
$\boldsymbol\psi=(0,0,\psi(x,y))$. If this is the case  (4) is reduced to a
linear PDE $$ \Delta\psi=\lambda\epsilon\psi, \eqno(5) $$ where $\Delta$ is a
two dimensional Laplassian.

Hence, studying the equations (1,5) is of interest for continuous media
electrodynamics. A similar case can be made in acoustics, whence comes the
original name of the equation (1). In both cases the equations describe
transmission of signals, and are quite relevant for applications.

For the Schr\"{o}dinger equation one of the most efficient ways of building
potentials allowing exact solutions, is the method of factorization, or the
Darboux transformation. Developing a similar formalism for the equation (1)
appears to be a natural thing to do. Its basics are presented in the next two
sections of this article. In particular, we derive the related generalized Crum
formulae, build chains of discrete symmetries and study their simple closing.
We argue that there is no one-to-one correspondence between this formalism and
the well-known technique for the Schr\"{o}dinger equation [5,6,10].

The acoustic problem (1) is also interesting from the integrable systems viewpoint. It
is known that this equation represents the L-equation of the Lax pair (or the
LA-pair) for the nonlinear Harry Dym (HD) PDE: see [1-3] and references
therein. We illustrate the Darboux transformation technique for the ODE (1) by
constructing a new positon solution of the HD equation from a pair of its
simple soliton solutions.

The Darboux transformation is known not only as a way of finding exact
solutions of nonlinear equations, but also as a resource for proliferation of
these equations, that is building new integrable PDEs together with their Lax
pairs by means of the dressing chain technique. As an illustration, we
construct a modified Harry Dym (mHD) equation (and its Lax pair) which has a
remarkably simple form.

In the last section we turn to the equation (5). The Moutard transformations
for this equation provide a simple method for construction of exact solutions
of the Maxwell equations (2) with a dielectric permitivity
$\epsilon=\epsilon(x,y)$, which in general has a complicated singularity
structure. However, a simple periodic closing of a dressing chain, generated by
the Moutard transforms results in a regular integrable 2D dielectric
permitivity.

\head Discrete symmetries of the one-dimensional acoustic problem\endhead

The Darboux transformation technique makes use of the existence of specific
discrete isospectral symmetries of the equation under consideration and is
standard in the theory of integrable PDEs: e.g. [5,6] and references therein.

We start out with a standard non-linear substitution  [1,2,4] due to which
typically the solutions of (1) will be represented parametrically: $$
u(x)=v_y(y),\qquad x=v(y). \eqno(6) $$ This reduces the equation (1) to $$
\psi_{yy}=U\psi_y+\lambda\psi, \eqno(7) $$ with $U=v_{yy}/v_y$. The above
quantity $U(y)$ will be referred to as a potential as well as the dielectric
permitivity function $\epsilon(x)=\frac{1}{u^2(x)}$ mentioned earlier.

In spite of the fact that (7) is easily reducible to the Shr\"{o}dinger
operator, from the point of view of finding integrable potentials this
connection is not trivial. Let us address this issue in more detail.

A substitution $$ \psi\to \sqrt{v_y}\psi, $$ transforms (7) into the stationary
Schr\"odinger equation $$ \psi_{yy}=\left(\lambda+V(y)\right)\psi, \eqno(7.1)
$$ where the potential $V(y)$ is related to the potential $U(y)$ of the
acoustic problem (7) via $$ V=\frac{U^2-2U_y}{4}. \eqno(7.2) $$ Linearizing
(7.2) with a substitution $U=-2p_y/p$, one sees that $p(y)$ in turn satisfies
(7.1) with $\lambda=0$.

The Darboux transformation for (7.1) is well known and at the first sight it
may appear that developing an independent technique for the acoustic problem
(7) is superfluous. However, the following argument shows that this is not the
case. Namely, there is no one-to-one correspondence between the problems (7)
and (7.1).

Indeed, let $U(y)$ be a specific potential for the acoustic problem (7), not
depending on any free parameters. From (7.2) one can get (uniquely) the
Shr\"odinger potential $V(y)$ and further substitute it into (7.1). In order to
reconstruct the initial potential $U(y)$, equation (7.1) should be solved with
$\lambda=0$. Let the solution be $p=p(y,C_1,C_2)$, depending on a pair of
constants $C_1,C_2$. One of these constants, say $C_1$, plays the normalizing
role and can be omitted. Then the restored potential $U=U(y,C_2)$ will depend
not only on $y$, but the free parameter $C_2$ as well. Hence, a single
potential in the  Shr\"odinger operator generates the whole family of
potentials for the acoustic problem, and in order to single out a specific
potential for the latter one would have to subsequently develop some selection
mechanism by studying the sequence of maps $U(y)\rightarrow V(y)\rightarrow
U(y,C_2)\rightarrow U(y)$. This necessity gets bypassed if one develops the
Darboux transform formalism directly apropos of the operator (7) without using
(7.1), and this is done in this and the following section.

Following Shabat [5] we seek elementary discrete symmetries of the equation (7)
effecting the change $$ \psi\to \psi^{(1)}=f\psi_y+g\psi, \eqno(8) $$ for some
$\lambda$-independent functions $f$ and $g$ of $y$.

One easily verifies that there are three distinct discrete symmetries of the
type (8) for the equation (7). They are $$ \aligned \psi\to
\psi^{(1)}=\frac{\psi}{v},\qquad v\to v^{(1)}=\frac{1}{v};\qquad
\psi\to\psi^{(1)}=\frac{\psi_y}{v_y},\qquad v\to v^{(1)}=\int\frac{dy}{v_y}
\endaligned
\eqno(9) $$ and $$ \matrix
\psi\to\psi^{(1)}=\frac{\psi_1\psi_y}{\psi_{1,y}}-\psi,\cr \\ v_y\to
v^{(1)}_y=v_y\left(\frac{\psi_1}{\psi_{1,y}}\right)^2, \qquad U\to
U^{(1)}=U+2D\ln\frac{\psi_1}{\psi_{1,y}}.
\endmatrix
\eqno(10) $$ In the latter equation $\psi_1=\psi_1(y,\lambda_1)$ is a
particular solution of (7) with the spectral parameter value $\lambda_1$,
further referred to as a ``prop solution",  $D=\partial_y$ and
$\psi_{1,y}=D\psi_1$.

Note that the former two symmetries (9) define the new quantity
$U^{(1)}=\frac{v^{(1)}_{yy}}{v^{(1)}_y}$ in a way independent of any solution
$\psi(y,\lambda)$ of (7). These symmetries arise as a particular case of (8) as
the result of gauging corresponding to the choice of $f$ or $g$ alternatively
zero. These symmetries have a trivial kernel in the solution space of (7).
According to the terminology of [5] we call the symmetries (9) T-symmetries,
sometimes referred to as Schlesinger transforms\footnote{In the context of
soliton solutions, the T-symmetries play the part of explicitly invertible
B\"{a}cklund transforms [7].}.

On the other hand, the transformation (10) alias the Darboux transformation,
which [5] calls an S-symmetry, does have a non-trivial kernel on the solution
space of (7) (one can let $\psi=\psi_1$ in the first equation of (10) and get
zero). This property will be essential in the sequel.

Along the way, we shall use the popular term ``dressing" for the application
procedure of the transformation (10) to a triple $(\psi,v,U)$, the resulting
pair $(\psi^{(1)},v^{(1)},U^{(1)})$ being referred to as the  ``dressed" one.

Despite a nearly trivial countenance,  the Darboux transform (10) has a
remarkable capacity to enable one to engineer potentials with arbitrary
discrete spectra {\it ad hoc}. Indeed, suppose it is possible to solve the
equation (7) formally (namely, obtaining among others some ``non-physical"
solutions which are not in $L^2$) for some potential $U$ and all $\lambda\in
\Bbb R$. Suppose, $\psi_1(y,\lambda_1)$ is such a solution. Let us denote its
linearly independent counterpart as $\hat{\psi}_1(y,\lambda_1)$, i.e. $$
\hat\psi_1=\psi_1\int dy\frac{v_y}{\psi_1^2}. $$ Dressing $\hat\psi_1$
according to (10), we find $$ \hat\psi_1^{(1)}=\frac{v_y}{\psi_{1,y}}. $$

Therefore, if one comes up with a non-physical prop solution $\psi_1$ by
requiring that its derivative $\psi_{1,y}$ be strictly positive and rapidly
growing as $|y|\to\infty$, then in the spectrum of the dressed potential
$U^{(1)}$  there will appear a level $\lambda_1$, not present in the original
spectrum for $U$. Since the principle for the choice of the value of
$\lambda_1$ is such that this value is not to be present in the physical
spectrum for $U$, repeating the dressing procedure $n$ times will result in a
potential $U^{(n)}$ possessing $n$ new pre-chosen levels
$\lambda_j,\,j=1,\ldots,n$. See (13,14) in the sequel.

Conversely, the function $\hat\psi_1^{(1)}$ generates an inverse transformation
(undressing) to (10). Thus, one can as well remove some pre-chosen levels from
the spectrum of a potential.

\head Crum formulae and dressing chains for the classical acoustic spectral
problem \endhead

Below we present the formulae describing an $n$-step dressing procedure for any
$n\geq1$, whose analogues are known for the Schr\"{o}dinger equation as the
Crum formulae [8]. We derive them for equation (7) following the procedure
exposed in [9].

A single act of dressing (10) can be iterated $n$ times to yield a triple
$(\psi^{(n)},v^{(n)}, U^{(n)})$. One starts out by dressing a triple
$(\psi,v,U)\equiv(\psi^{(0)},v^{(0)},U^{(0)})$ corresponding to a spectral
parameter $\lambda$ with a prop function $\psi_1\equiv\psi_1^{(0)},$ which is a
formal solution of  (7) with a spectral parameter $\lambda_1$ and the potential
$U^{(0)}$. The resulting solution $\psi^{(1)}$ solves (7) with the dressed
potential $U^{(1)}$ (and the same spectral parameter $\lambda$). On the $j$th
step, $j=1,\ldots,n$ one uses some prop solution $\psi_j^{(j-1)}$ which solves
(7) with a pre-dressed potential $U^{(j-1)}$ and a spectral parameter value
$\lambda_j$ to produce the $j$ times dressed solution $\psi^{(j)}$ and
potential $U^{(j)}$ (as well as the function $v^{(j)}$ with
$U^{(j)}=v^{(j)}_{yy}/v^{(j)}_y$). Note that the spectral parameter $\lambda$
in the dressed equations for $\psi^{(j)}$ is the same for all $j=1,\ldots,n$.

It's easy to see that the $n$ times dressed solution $\psi^{(n)}$ shall have
the form $$ \psi^{(n)}=\sum_{j=1}^n a_jD^j\psi+(-1)^{n}\psi, \eqno(11) $$ with
the  functions-coefficients $a_j$ to be found, which of course will depend on
the choice of the prop solutions $\psi_j^{(j-1)}$. It follows from (10) that $$
U^{(n)}=U+2D\ln a_n, \eqno(12) $$ for $$ \psi^{(n)}= \prod_{j=1}^n
\frac{\psi_j^{(j-1})}{D\psi_j^{(j-1)}} D^n\psi^{(0)}\, +\, \ldots\, +  \,
(-1)^n\psi^{(0)}, \qquad U^{(n)}=U^{(0)}+2D\ln\prod_{j=1}^n
\frac{\psi_j^{(j-1})}{D\psi_j^{(j-1)}}, $$ where the ellipses in the first
formula stand for the terms containing the derivatives of $\psi^{(0)}$ of
orders  from $1$ through $n-1$.

So far the choice of the prop solutions $\psi_i^{(i-1)}$ has been quite
arbitrary. But suppose now that the original equation (7) possesses $n$
distinct formal solutions $\psi_j$, corresponding to spectral parameter values
$\lambda_j, \, j=1,\ldots,n$. Let $\psi_j\equiv\psi^{(0)}_j$ and consider the
following dressing procedure (which will be further used for the dressing chain
construction):

$$ \matrix \psi^{(0)}  & &\boldsymbol \psi_{\boldkey 1}^{\boldkey (\boldkey 0
\boldkey )} & \psi_2^{(0)} &\ldots& \psi_{n-1}^{(0)} & \psi_n^{(0)}&& U^{(0)}
\cr \psi^{(1)} && 0 & \boldsymbol \psi_{\boldkey 2}^{\boldkey (\boldkey 1
\boldkey )} &\ldots& \psi_{n-1}^{(1)} & \psi_n^{(1)}&& U^{(1)} \cr \vdots & & &
& & & && \vdots \cr \psi^{(n-1)} && 0 & 0& \ldots & 0 & \boldsymbol
\psi_{\boldkey n}^{\boldkey (\boldkey n\boldkey -\boldkey1 \boldkey )} &&
U^{(n-1)} \cr \psi^{(n)} && 0 & 0& \ldots &0& 0 && U^{(n)}
\endmatrix
\eqno(\boldkey *) $$ Namely, for $j=1,\ldots,n$ on the above diagram ($\boldkey
*$) every new line $j+1$ is obtained by dressing the functions from the
preceding line $j$ by (10) with a prop solution $\boldsymbol \psi_{\boldkey
j}^{\boldkey (\boldkey j\boldkey -\boldkey1 \boldkey )}$ marked in {\bf bold}.

Zeroes, proliferating as one moves down the diagram stem from the non-trivial
kernel property of the S-symmetry, and it is this property that now enables one
to find the unknown functions $a_j$. Indeed, substitution of any
$\psi_j=\psi^{(0)}_j$ for $\psi$ in the right hand side of (11) shall yield
zero. Hence, the coefficients $a_j$ satisfy a system of $n$ independent linear
algebraic equations, namely $$ \sum_{k=1}^n a_kD^k\psi_j+(-1)^n\psi_j=0,\qquad
j=1,..,n. $$ Solving it by the Kramer rule and substituting the result into
(11) and (12), we end up having: $$
U^{(n)}=U+2D\ln\frac{\widetilde\Delta_n}{\Delta_n},\qquad \text{i.e.}\qquad
v_y^{(n)}=v_y\left(\frac{\widetilde\Delta_n}{\Delta_n}\right)^2,
\qquad\psi^{(n)}=\frac{\widehat\Delta_{n+1}}{\Delta_n}, \eqno(13) $$ where
$\Delta_n$, $\widetilde\Delta_n$ are determinants of square $n\times n$
matrices, whereas $\widehat\Delta_{n+1}$ -- of an $n+1\times n+1$ matrix as
follows: $$ \Delta_n=\vmatrix D\psi_1&\ldots&D^n\psi_1 \cr \vdots & & \vdots
\cr D\psi_n&\ldots&D^n\psi_n
\endvmatrix,
\qquad \widetilde\Delta_n=\vmatrix \psi_1&\ldots&D^{n-1}\psi_1  \cr \vdots & &
\vdots  \cr \psi_n&\ldots&D^{n-1}\psi_n
\endvmatrix,
$$
\newline
$$ \widehat\Delta_{n+1}=\vmatrix \psi &\ldots&D^{n}\psi \cr
\psi_1&\ldots&D^n\psi_1 \cr \vdots && \vdots  \cr \psi_n&...&D^n\psi_n
\endvmatrix.
\eqno(14) $$ Note that the linearity of (7) makes the choice of the sign before
$\psi^{(n)}$ irrelevant.

The obtained formulae (13,14) make it possible to find rich families of exact
solutions of (1). The easiest case is dressing from the birthday suit, or on
the vacuum background, assuming $U=0$ (hence $x=c_1y+c_2$, $u(x)=c_1$, where
$c_1$ and $c_2$ are arbitrary constants).

Such a natural rigging yields gratis all the B-potentials reported in [4],
moreover the formulae for their computation derived therein turn out to be
particular cases of (13,14) with merely $v_y=1$. For instance
$\psi_1=\sinh\xi_{1}$, $\psi_2=\cosh\xi_{2}$  (here $n=2$, $y_{j}$ are
constants, $\xi_{i}=k_{j}(y-y_{j})$, $j=1,2$) yield a reflexionless B-potential
with a power $2/3$ singularity: $\epsilon^{(2)}=\frac{1}{[u^{(2)}]^2}$,
where\footnote{Note that the expressions for $u^{(2)}$ are parametric. In order
to interpret the formulae correctly, the reader is referred back to (1,6,7).
The orders of the singularities pertain to the potential $\epsilon(x)$, which
is a zero of the function $u(x)$ and a singularity of the potential
$U(y)=v_{yy}/v_y$, where the function $v(y)$ solves the equation
$u[v(y)]=v_y(y)$.} $$
u^{(2)}=\left(\frac{k_{2}\sinh\xi_{1}\sinh\xi_{2}-k_{1}\cosh\xi_{1}\cosh\xi_{2}}
{k_{1}\sinh\xi_{1}\sinh\xi_{2}-k_{2}\cosh\xi_{1}\cosh\xi_{2}}\right)^2. $$ The
same $\psi_1$ and $\psi_2=\sinh\xi_{2}$ yield another potential with a power
$4/5$ singularity: $$
u^{(2)}=\left(\frac{k_{2}\sinh\xi_{1}\cosh\xi_{2}-k_{1}\sinh\xi_{2}\cosh\xi_{1}}
{k_{1}\sinh\xi_{1}\cosh\xi_{2}-k_{2}\sinh\xi_{2}\cosh\xi_{1}}\right)^2. $$ By
construction, these potentials have only two levels $\lambda_{1,2}$.

In addition, all the {\it regular} reflexionless potentials can be also built
by the formulae (13,14) once again by dressing $U=0$. Here comes the proof. The
passage coefficient for a regular $n$-level reflexionless potential can be
expressed by a well-known formula $$
T_n(p)=\prod_{j=1}^{n}\frac{k_j-ip}{k_j+ip}, $$ with $\lambda_j=k_j^2$. As has
been pointed out earlier, the levels $\lambda_j$ can be successively removed
from the spectrum by means of the inverse of the Darboux transformation (still
having the form (10)), each application of which will kill a term in the
product. Successively applying this procedure $n$ times, for the passage
coefficients we have $$ T_n(p)\to T_{n-1}(p)\to...\to T_0(p)=1. $$ This proves
our assertion, because the case $R(p)=0$ and $T(p)=1$ for the reflection and
the passage coefficient for all $p$ is feasible with $U=0$ only. It's worth
reiterating the point that the argument above owes itself to the fact that the
S-symmetry possesses a non-trivial kernel in the space of solutions of (7).

Matveev and Salle [6] find super-reflexionless potentials for the KdV equation,
alias positons. In the same vein one can operate on the equations (1,7). In
order to do so, one should use the formulae (13,14) with  $n=2$ choosing the
prop solutions $\psi_{1,2}$ respectively as $\psi_1(y,\lambda_1)$ and
$\psi_1(y,\lambda_1+\delta),$ and then letting $\delta\to 0$. If withal $U=0$
and $\psi_1$ generates a single soliton potential, then (13) defines a single
positon potential. The next section describing the positon solutions of the HD
equation contains the aforementioned computation.

In addition to B-potentials, various other interesting ones can be produced.
For instance, one can construct soluble potentials with a finite equidistant
spectrum\footnote{The same statement applies to the (stationary)
Schr\"{o}dinger equation.}.

One can also investigate potentials which change in a specific simple way under
the Darboux transform, e.g. such that $U\to U+$const or $U\to$const$U$. For the
Schr\"{o}dinger equation, the former transformation is shape-invariant and for
$n=1$ results in the harmonic oscillator potential. Let's develop an analogue
for the model (1,7) under investigation. Suppose, $$
U^{(1)}=U+\frac{2}{\omega^2}, $$ for a constant $\omega$. Then we can obtain
parametrically the function $u(x)$ from (1) as follows: $$ u(x)=\alpha
z\exp\left(-\omega^2z-\frac{\kappa^2}{z}\right),\qquad x=x_0-\alpha
\omega^2\int dz \exp\left(-\omega^2z-\frac{\kappa^2}{z}\right), $$ where,
$\kappa$, $x_0$, $\alpha$ are real constants and $z=\exp(-y/\omega^2)$. The
prop function $\psi_1$ rendering the potential $U^{(1)}$ from $U$ has the
countenance $\psi_1=\exp(-\omega^2z)$ and solves (7) with an eigenvalue
$\lambda_1=b^2/\omega^2$. It's easy to verify that the dielectric permitivity
$\epsilon(x)=1/u^2(x)$ has a second order pole at $x=x_0$.

The theory of the Darboux transformation for the Schr\"{o}dinger equation
utilizes the concept of dressing chains of discrete symmetries and their
closing. The work of Veselov and Shabat [10] elucidates how the dressing chain
closing method can be used in order to obtain various potentials with
meaningful mathematical physics. Namely, a simple closing procedure leads one
to the harmonic oscillator potential (resulting also in a shape-invariant
change of potential). A more complicated closing results in finite-gap
potentials as well as the fourth and the fifth Peinleve equations, see [10].

Dressing chains can be written out for the equation (7) as well. Let us
introduce a sequence $\{f_n\}_{n\geq1}$ of functions as follows: $$
f_n=D\ln\psi_n^{(n-1)}, $$ with the quantity $\psi_n^{(n-1)}$ as it has been
introduced in the diagram ($\boldkey *$) above (where it appeared in {\bf
bold}). In particular, it corresponds to the pre-chosen value $\lambda_n$ of
the spectral parameter.

One can verify by hand starting from $n=1$ that $$ U^{(n)}=U-2D\ln\prod_{j=1}^n
f_j. $$ Besides, direct substitution shows that $f_n$ satisfies the equation $$
f_n'+f_n^2-U^{(n)}f_n=\lambda_n, $$ where $f'=Df$. The two latter relations
imply the recursion connecting $f_n$ and $f_{n+1}$ as follows: $$ (f_n
f_{n+1})'=f_n f_{n+1} (f_n - f_{n+1} ) + \lambda_{n+1}f_n - \lambda_n f_{n+1}.
\eqno(15) $$ This equation (15) represents a dressing chain for the acoustic
problem.

In a way analogous to the theory of dressing chains for the Schr\"{o}dinger
equation [10] we are interested in T-periodic chain closing, namely  imposing
the condition $f_{n+T}=f_n$ for an integer $T\geq1$. We shall consider here the
easiest case $T=1$.

Given the spectral parameter values $\lambda_{1,2}$, one obtains a
one-parameter family of potentials, indexed by a constant $c$: $$
U=\frac{-(\lambda_1-\lambda_2)^2y^2 +
2c(\lambda_2-\lambda_1)y+6\lambda_1-2\lambda_2-c^2} {2[(\lambda_1-\lambda_2)y
+c]}. $$ If $\lambda_2=3\lambda_1>0$ and $c=0$, we can express the function
$u(x)$ parametrically: $$ x(y)=\frac{\sqrt{\pi}}{2\alpha} \text{Erf}(\alpha
y),\qquad u(y)=\exp\left(-\alpha^2 y\right), $$ where $\alpha^2=-\lambda/2>0$.

It is known that for the Schr\"{o}dinger equation, a nontrivial chain closing
operation with $T>1$ results in  finite gap potentials [10]. Such potentials
for the HD equation are due to Dmitrieva [3]. A close analogy between the
Schr\"{o}dinger equation and the acoustic problem  [1,2,11] suggests that one
can expect results similar to those of [10] apropos of the analysis of higher
order chain closing for $T>1$. We expect that potentials built in such a way
can have interesting physical applications, such as for instance a model of
wave propagation in media whose dielectric permitivity is a periodic function
of a single spatial variable.

\head HD and mHD equations \endhead

The 1+1 HD equation $$ u_t=u^3u_{xxx}+\beta u_x, \eqno(16) $$ with some real
constant $\beta$, has been studied quite extensively since late 70's: see [1-3]
and references therein. It arises in the study of evolution equations solvable
via the spectral transforms method based on the string rather than the
Schr\"{o}dinger equation. The principal approach to it has been based on its
relation to the  KdV, mKdV and other more classical hierarchies of integrable
PDEs [2,3]. However, as was shown above, this relation is not entirely
straightforward, and the direct approach developed herein enables one to
produce new solutions of the HD equation in addition to those already known. As
an example, below we construct a simple positon solution.

The acoustic problem (1) is the first equation in the LA-pair for the HD
equation (16), the full pair being $$ \cases
\psi_{xx}=\frac{\lambda}{u^2}\psi,\\ \hfill \\ \psi_t = \left(4\lambda
u+\beta\right)\psi_x-2\lambda u_x\psi.
\endcases
\eqno(17) $$ The coordinate change (6) in the presence of time dependence
becomes $$ t\to t,\qquad x\to v(y,t), $$ thus $$
\partial_x\to \frac{1}{v_y}\partial_y,\qquad
\partial_t\to\partial_t-\frac{v_t}{v_y}\partial_y.
$$ After this change (17) becomes $$ \cases
\psi_{yy}=\frac{v_{yy}}{v_y}\psi_y+\lambda\psi,\\ \hfill \\
\psi_t=\left(\frac{v_t+\beta}{v_y}+4\lambda\right)\psi_y- \frac{2\lambda
v_{yy}}{v_y}\psi,
\endcases
\eqno(18) $$ and the HD equation (16) transforms to $$
v_y(v_tv_{yy}-v_{yt}v_y)+3v_{yy}^2+ v_y(v_{4y}v_y-4v_{3y}v_{yy}+\beta
v_{yy})=0. \eqno(19) $$ with the notations $v_{3y}$, $v_{4y}$ for the partial
derivatives in $y$ of order 3 and 4 respectively.

The goal now is to extend the Darboux transformation (10) for the equation (7)
alias the first equation in (18), so that it agrees with the second equation in
the Lax pair. One just includes the $t$-dependencies in (10). At this point it
only provides the  value of the partial derivative $$
v^{(1)}_y=\left(\frac{\psi_1}{\psi_{1,y}}\right)^2\equiv A(y,t), $$ rather than
the dressed quantity $v^{(1)}(y,t)$ of interest. Hence, let $v^{(1)}_t=B(y,t)$
be unknown and let's assume that $v^{(1)}$ satisfies the second equation of the
pair (18)  with the dressed according to (10) function $\psi^{(1)}$, the
quantities $(\lambda,\beta)$ remaining the same. One can express the unknown
quantity $B$ as follows: $$ B=\left(\frac{\psi_1}{\psi_{1,y}}\right)^2\left(
\beta+4\mu v_y+v_t-2v_yU_y\right)+ \frac{4\psi_1
v_{yy}}{\psi_{1,y}}-4v_y-\beta, $$ and verify that $B_y=A_t$. It follows that
$$ v^{(1)}(y,t)=\int Ady+Bdt, \eqno(20) $$ with a closed 1-form under the
integral.

Hence, the Darboux transformation (10,20) is an S-symmetry for the LA-pair
(18), and therefore for the HD equation (19).

This enables one to construct exact solutions for this equation. We exemplify
it with a single soliton solution and a single positon solution.

Let $v=y$, $\lambda_1=k^2$, $\psi_1(y,t,k)=\sinh[\phi(y,t,k)]$ with
$\phi=k\left(y+(4k^2+\beta)t\right)$, then by (20) one has $$
v^{(1)}=\frac{1}{k^3}\left(\phi-\tanh \phi\right)-(4+\beta)t. $$ The function
$v^{(1)}(y,t,k)$ determines a single soliton B-potential
$U^{(1)}=v^{(1)}_{yy}/v^{(1)}_y$, mentioned in the previous section.

The single positon potential is obtained from two distinct soliton solutions
$\psi_1(y,t,k)$ and $\psi_1(y,t,k+\delta )$, using them as the prop functions
$\psi_{1,2}$  in the formulae (13,14) with $n=2$ and taking the limit as
$\delta \to 0$. Namely, $$
v^{(2)}_y=v_y\left(\frac{\psi_{1,yyk}\psi_{1,y}-\psi_{1,yy}\psi_{1,yk}}
{\psi_{1,yk}\psi_1-\psi_{1,y}\psi_{1,k}}\right)^2, $$ where the subscript
${}_k$ means differentiation by $k$.  Taking $\psi_1$ explicitly as the
hyperbolic sine in the previous example results in $$
v^{(2)}_y=k^4\left(\frac{\sinh(2\phi)+2\tilde\phi}{\sinh(2\phi)- 2\tilde
\phi}\right)^2, $$ with $\tilde\phi=k\left(y+(12k^2+\beta)t\right)$.

It is well known that the dressing formalism enables one to produce hierarchies
of integrable PDEs. Borisov and Zykov [12] proposed a technique for
proliferation of integrable equations, which they applied to the KdV and the
Sine-Gordon (SG) equations. The technique is based on the discrete symmetries
dressing chain closing. The main idea of the approach is as follows. The
equation (for illustration purposes let us take the KdV equation) is written as
a compatibility condition of a pair of equations, further denoted as $L_1$ and
$A_1$. Each of these equations is quadratic in the auxiliary field. Using
invariance of the pair with respect to the Darboux transformation (which is
viewed as a discrete symmetry), a second pair $L_2$, $A_2$ of equations is
built. Excluding the potentials from $L_1$, $L_2$ and $A_1$, $A_2$, it is
possible to obtain two equations, which [12] calls an $x$ and a $t$ chain,
respectively. (We further use the notations $C_x$ and $C_t$ instead.) If a
potential is excluded from $L_1$ and $A_1$, one ends up with a modified
equation mKdV. The equations $C_x$ and $C_t$ can be converted into the Lax pair
for the  mKdV equation in two ways, the Darboux transformation being already
known.

This procedure can be repeated, producing new equations with their LA-pairs. In
this vein, the equations m${}^2$KdV and m${}^3$KdV were obtained. The former
becomes the exponential Calogero-Degasperis  equation [11] after an exponential
change, the latter contains an elliptic equation of the same authors.

In spite of its simplicity, the technique described is very powerful. This can
be illustrated by the following examples. First, see  [13], the m${}^N$KdV
equations with $N=0,...,3$ together with the Krichever-Novikov equation
exhaust (modulo a contact transformation) all the integrable equations of the
form $u_t+u_{xxx}+f\left(u_{xx},u_x,u\right)=0$. Second, applying their
approach to the SG equation, the authors of [12] have succeeded to come up with
a new (!) nonlinear equation already on the second step. This equation has a
non-trivial B\"{a}cklund transform, admitting an interesting $2\pi$-kink-shelf
solution.

The same technique was shown to be applicable to the study of considerably more
difficult (1+2)-dimensional nonlinear PDEs. For instance, in [14] the
proliferation procedure was successfully adapted to the  Kadomtsev-Petviashvili
and Boiti-Leon-Pempinelli equations.

Let us apply this formalism to the HD equation. First note that the LA-pair for
(19) can be written as a system of two Ricatti equations: $$ \cases
g_y=-\lambda g^2-Ug+1,\\ \hfill\\
g_t=\lambda\left(2U_y-\frac{v_t+\beta}{v_y}-4\lambda\right)g^2-
\left(\frac{v_{ty}}{v_y}+4\lambda U\right)g+\frac{v_t+\beta}{v_y}+4\lambda.
\endcases
\eqno(21) $$ The second summand in the right hand side of the second equation
has a term, denoted as $v_{ty}$, representing a fairly long expression which
can be derived from (19). The function $g=g(y,t)$ is connected with the
solution $\psi$ of (18) as $g=\psi_y/\psi$. Excluding the function $v$ from
(1.1) and returning to the old variables via $x=g$, $u=g_y$, we obtain a
modified Harry Dym (mHD) equation: $$
u_t=u^3u_{3x}+3u^2u_xu_{xx}-3\lambda^2xu^2-\frac{3u^2(uu_{xx}+u_x^2)}{x}+
\frac{6u^3u_x}{x^2}+\frac{3u^2(1-u^2)}{x^3}. \eqno(22) $$ (By analogy with the
equation mKdV in [12], we call (1.2) the mHD equation.) As one can see, this
equation has a different countenance than the HD equation. However, omission of
all the summands but the first one in the right hand side of (22), yields the
HD equation (16) with $\beta=0$. Note that (22) can be rewritten quite nicely
in new variables  $x=1/z$, $u(x,t)=\sqrt{\theta(z,t)}$: $$
\left(\theta^{-1/2}\right)_t=\frac{3\lambda^2}{z}+
z^3\left(\frac{1}{2}z^3\theta_{3z}+\frac{9}{2}z^2\theta_{zz}+
9z\theta_z+3\theta-3\right). \eqno(23) $$ The formula (23) can be simplified
even further by changing $$ \theta=e^{-2\xi}\eta(\xi,t)+1,\qquad y=\log z. $$
As the result, it becomes $$
\left[\left(e^{-2\xi}\eta+1\right)^{-1/2}\right]_t=3\lambda^2 e^{-\xi}+
\frac{1}{2}e^{\xi}\left(\eta_{_{3\xi}}-\eta_{_{\xi}}\right). $$ However, we
will be considering the mHD equation in the form (23). Note than in the
stationary $\theta_t=0$ case, it reduces to a linear ODE!

The dressing chain method produces not only the equation (23), but also its
LA-pair. It is constructed as follows. Return to the chain (15) and let
$f_n=1/g$, $f_{n+1}=\Psi$, $\lambda_n=\lambda$, $\lambda_{n+1}=\mu$.
Considering $\mu$ as a spectral parameter, one can see that (15) can be viewed
as an L-equation of the LA-pair for equation (23). One should also define the
second, non-stationary chain $C_t$ for the functions $g_n$ (in terms of the
dynamical equation in (18)) and build the A-equation. Omitting the lengthy but
straightforward computation, we present the LA-pair for equation (23), written
in the variables $t$, $z$: $$ \cases \Psi_z=\frac{\mu}{z^2\sqrt{\theta}}\Psi^2+
\left(\frac{1}{z}+\frac{1}{z\sqrt{\theta}}-\frac{\mu}{z^3\sqrt{\theta}}\right)\Psi-
\frac{1}{z^2\sqrt{\theta}},\\ \hfill\\ \Psi_t=\mu a\Psi^2+b\Psi+c,
\endcases
\eqno(24) $$ where $$ \matrix a&=&
-4\mu+2\lambda-\frac{\lambda^2}{z^2}-2\lambda\sqrt{\theta}+
\left(\theta-1\right)z^2+2z^3\theta_z+\frac{1}{2}z^4\theta_{zz}, \\ \cr
b&=&4\left(\frac{\lambda}{z}-z-z\sqrt{\theta}\right)\mu+\frac{1}{2}z^5\theta_{zz}+
2z^4\theta_z+\left(\theta-\frac{\lambda}{2}\theta_{zz}-1\right)z^3 \\ \cr &-&
3\lambda z^2\theta_z+3\lambda(1-\theta)z-\frac{3\lambda^2}{z}+
\left(\frac{\lambda}{z}\right)^3,\\ \cr
c&=&4\mu-\frac{1}{2}z^4\theta_{zz}-3z^3\theta_z+\left(1-3\theta-
2\sqrt{\theta}\right)z^2-2\lambda+\left(\frac{\lambda}{z}\right)^2.
\endmatrix
$$ Note that the spectral parameter in (24) is $\mu$, whereas $\lambda$ enters
the non-linear equation (23). As one can see, the LA-pair for (23) also has the
form of a pair of Ricatti equations. These equations can be simultaneously
linearized in order to represent (23) as a compatibility condition for two
linear equations, as it is done in the theory of solitons.

We will not proceed further with the mHD equation. To conclude this section, we
would like to repeat the statement that the exact solutions of (23) are easily
found via the dressing technique, and this procedure can be extended in order
to produce the mHD equation and its LA-pair.  It is worth emphasizing that the
equations HD and (23) are members of different hierarchies. Thus, discrete
symmetries enable one to establish connections between different integrable
equation hierarchies, promoting the unification of knowledge about them.

\head Moutard transformations \endhead

We devote this last section to equation (5) which has been obtained from the
Maxwell equations (2) in the case of an isotropic but inhomogeneous in two
directions $(x,y)$ medium.

Clearly, a PDE (5) is harder to investigate than an ODE (1). Nevertheless, its
analysis in terms of the Darboux transform (10) known also as the Moutard
transformation [6] is quite similar to its ODE cousin. Below we shall present
the relevant formulae without the derivation details.

Let $\psi=\psi(x,y)$ and $\phi=\phi(x,y)$ be two particular solutions of (5),
namely $$ \Delta\psi-\lambda\epsilon\psi= \Delta\phi-\lambda\epsilon\phi=0.
\eqno(25) $$ We choose the function $\phi$ as a prop solution. Then the
following transformations represent an analogue of (10): $$
\psi\to\psi^{(1)}=\frac{\theta[\psi,\phi]}{\phi},\qquad \epsilon\to
\epsilon^{(1)}=\epsilon-2\lambda\Delta\ln\phi, \eqno(26) $$ where $$
\theta[\psi,\phi]=\int_{\Gamma}dx_{\mu}\varepsilon_{\mu\nu}
\left(\phi\partial_{\nu}\psi-\psi\partial_{\nu}\phi\right). \eqno(27) $$ Above,
the following (standard) tensor notations have been used: $\mu\in\{1,2\}$,
$x_{\mu}\in\{x,y\}$, $\partial_{\mu}=\partial/\partial x_{\mu}$,
$\varepsilon_{\mu\nu}$ is a fully antisymmetric tensor with
$\varepsilon_{12}=1$, summation is implied over repeated indices. Note that a
one-form, which is being integrated in the formula (27) is closed in the case
when $\psi$ and $\phi$ are solutions of (26). Hence, the shape of the contour
of integration $\Gamma$ in (27) is irrelevant.

One can verify by direct substitution of (26,27) into the formulae (28) below,
that the dressed function $\psi^{(1)}$ satisfies the dressed equation (25)
(with the potential $\epsilon^{(1)}(x,y)$ and the same spectral parameter value
$\lambda$).

The Moutard transformation (26) can be iterated several times, and the result
can be expressed via Pfaffian forms [6]. Instead, we direct our interest to the
Maxwell equations (2). A straightforward computation (recall,
$\lambda=-{c^2\over\omega^2}$) yields the expressions for the dressed electric and
magnetic fields $\bold E^{(1)}$, $\bold B^{(1)}$: $$ {\bold E^{(1)}}=e^{i\omega
t}\left(0,0,\psi^{(1)}\right),\qquad {\bold B^{(1)}}=\frac{c}{\omega}e^{i\omega
t} \left(-\psi^{(1)}_y,\psi^{(1)}_x,0\right),\qquad {\bold
D^{(1)}}=\epsilon^{(1)}{\bold E^{(1)}}. \eqno(28) $$ On the basis of (28), one
can build a variety of exact solutions of the Maxwell equations. As a simple
example let's dress $\epsilon=0$. This isn't quite a medium, but one can easily
proceed with formal calculations (26-28) which result in a new ``medium" whose
dielectric permitivity $\epsilon^{(1)}(x,y)$ and the stationary component of
the field $\psi^{(1)}$ are as follows: $$
\epsilon^{(1)}=-\frac{8c^2}{\omega^2}\frac{a'(z)b'(\bar z)} {\left(a(z)+b(\bar
z)\right)^2},\qquad \psi^{(1)}=\frac{a(z)\beta(\bar z)-\alpha(z)b(\bar
z)+\xi(z,\bar z)} {a(z)+b(\bar z)}, \eqno(29) $$ where $$ \xi(z,\bar z)=\int
dz\left(\alpha(z)a'(z)-a(z)\alpha'(z)\right)+ \int d{\bar z}\left(\beta'(\bar
z)b(\bar z)-b'(\bar z)\beta(\bar z)\right), $$ $a(z)$, $\alpha(z)$, $b(\bar
z)$, $\beta(\bar z)$ are arbitrary functions of $z=x+iy$, ${\bar z}=x-iy$. Note
that the function $\psi^{(1)}$ from (26,28) provides in fact a general solution
of the dressed equation, for it is described in terms of two arbitrary
functions $\alpha(z)$ and $\beta(\bar z)$. To ensure that the quantities found
correspond to a physical non-absorbing medium, one should require that the
dressed dielectric permitivity function $\epsilon^{(1)}$ be real. This imposes
an extra restriction to the quantities $a(z)$ and $b(\bar z)$, namely $b(\bar
z)=\overline{a(z)}$. Generally speaking, the functions $\epsilon^{(1)}$ and
$\psi^{(1)}$ from (29) will have singularities along certain curves in the
$(x,y)$-plane.

The reflexionless B-potentials for the one-dimensional problem (1) above,
possess point singularities on the real line (corresponding to zeroes of the
function $u(x)$). Clearly, their 2D-analogues, such as (29) for the equation
(5) allow a much more diverse structure of singularities on the real plane. On
the other hand, not requiring that the quantity $\epsilon^{(1)}$ be real, one
obtains an absorbing medium which may not be devoid of interest for physical
applications.

In conclusion, let us study a dressing chain generated by the Moutard
transformations (26). A simple periodic closing of the dressing chain results
in a regular dielectric permitivity, similar to the 1D case studied above.

Denote $f_n=\ln\phi$, $f_{n+1}=\ln\psi^{(1)}$. Then after a straightforward
computation $$ \Delta\left(f_n+f_{n+1}\right)= \|\nabla f_n\|^2-\|\nabla
f_{n+1}\|^2, \eqno(30) $$ where $\|\cdot\|$ is the Euclidean norm.

The chain (30) is closely related to that of Veselov and Shabat [10] for the
Schr\"{o}dinger equation. Choosing $f_n$ specifically as $$
f_n=\sqrt{\lambda_n}y+\int dx\, g_n(x), $$ and substituting it into (30)
($\lambda_n$ being constant), we obtain for the quantities $g_n(x)$ the
following expression $$
\left(g_n+g_{n+1}\right)'=g_n^2-g_{n+1}^2+\lambda_n-\lambda_{n+1}, $$ matching
the corresponding formula of [10].

The simplest periodic closing of the dressing chain (30) is
$f_{n+1}=f_n=F(x,y)$, which implies that the latter function $F$ is harmonic,
and that the regular dielectric permitivity function in the corresponding
medium is given by the formula $$
\epsilon(x,y)=\frac{c^2}{\omega^2}\left(F_x^2+F_y^2\right). $$


\newpage
\Refs \ref\no 1 \by W. Heremah, P.P. Banerjee and M.R. Chatterjee \jour J.
Phys. A {\bf 22}\yr 1989\pages 241
\endref
\ref\no 2 \by L.A. Dmitrieva \jour J. Phys. A {\bf 26}\yr 1993\pages 6005
\endref
\ref\no 3 \by L.A. Dmitrieva \jour Phys. Lett. A {\bf 182}\yr 1993\pages 65
\endref
\ref \no 4 \by  V.S. Novikov \jour  JETP Lett. {\bf 72}\yr 2000\pages 223
\endref
\ref \no 5 \by  A.B. Shabat\yr 1999\jour Theor. and Math. Phys. V. 121. \pages
165
\endref
\ref \no 6 \by V.B. Matveev and M.A. Salle \book ``Darboux Transformation and
Solitons", Berlin--Heidelberg: Springer Verlag\yr 1991
\endref
\ref \no 7 \by  S.I. Svinolupov and R.I. Yamilov\yr 1994\jour Theor. and Math.
Phys. V.98. \pages 207
\endref
\ref \no 8 \by M M Crum\yr 1955\jour Quart. J. Math. Oxford, {\bf 6}, 2\pages
121
\endref
\ref \no 9 \by A.V. Yurov \jour Phys. Lett. A {\bf 225}\yr 1997\pages 51
\endref
\ref \no 10 \by A.P. Veselov and A.B. Shabat \jour Functional Anal. Appl. {\bf
27} no. 2 \yr 1993\pages 81
\endref
\ref\no 11 \by F. Calogero and  A. Degasperis \book ``Spectral Transform and
Solitons", North-Holland Publishing Company \yr 1982
\endref
\ref\no 12 \by B.A. Borisov and S.A.Zykov \jour Theor. Math. Phys., {\bf 115}
\yr 1998 \pages 530
\endref
\ref\no 13 \by A.V. Mikhailov, A.B. Shabat and R.I. Yamilov \jour Russ. Math.
Surv., {\bf 42} \yr 1987 \pages1
\endref

\ref\no 14 \by A.V. Yurov \jour Theor. Math. Phys. {\bf 119} \yr 1999 \pages
731
\endref
\endRefs
\enddocument